\documentclass[aps,prl,twocolumn,groupedaddress]{revtex4}

\usepackage{graphics}
\usepackage{epsfig}

\begin{document}

\title{Critical Behaviour in the Relaminarisation of Localised
Turbulence in Pipe Flow
}

\author{Ashley P. Willis}   
\email{A.Willis@bris.ac.uk}
\author{Rich R. Kerswell}
\email{R.R.Kerswell@bris.ac.uk}

\affiliation{Department of Mathematics, University of Bristol, University Walk,
   Bristol BS8 1TW, United Kingdom}

\date{\today}

\begin{abstract}
\end{abstract}


\noindent
{\bf Willis and Kerswell Reply:}
In our letter \cite{WK07}
it was reported that in pipe flow the median time $\tau$ for 
relaminarisation of localised turbulent disturbances closely follows the scaling
$\tau\sim 1/(Re_c-Re)$.
This conclusion was based on data from collections of 40 to 60 independent 
simulations at each of six different Reynolds numbers, $Re$.
In the previous comment, Hof {\it et al.} estimate $\tau$ differently for the 
point at lowest $Re$.  Although this point is the most uncertain, it forms the basis
for their assertion that the data might then fit an exponential scaling 
$\tau\sim \exp(A\,Re)$, supporting \cite{hof06}, for some constant $A$.  
The most certain point (at largest $Re$)
does not fit their conclusion and is rejected.
We clarify why their argument for rejecting this point is flawed.
The median $\tau$ is estimated from the distribution of observations, and
it is shown that the correct part of the distribution is used. 
The data is sufficiently well determined to show that
the exponential scaling cannot be fit to the data over this range of 
$Re$, whereas the $\tau\sim 1/(Re_c-Re)$ fit is excellent, 
indicating critical behaviour and supporting experiments \cite{peixinho06}.

The data median was used in our original analysis
as the estimator for the parameter of the exponential distribution.  Estimating via
the log-probability plot as in \cite{hof06}, the range of possible lines that can be 
drawn does not at all reflect the true error in the mean.
In the following we estimate confidence intervals by using the bootstrapping method 
\cite{effron79} applied to the maximum likelihood estimator of $1/\tau$.
Any truncated data is integrated in a logical fashion into this resampling method by
further resampling.

\begin{figure}[htb]
   \epsfig{figure=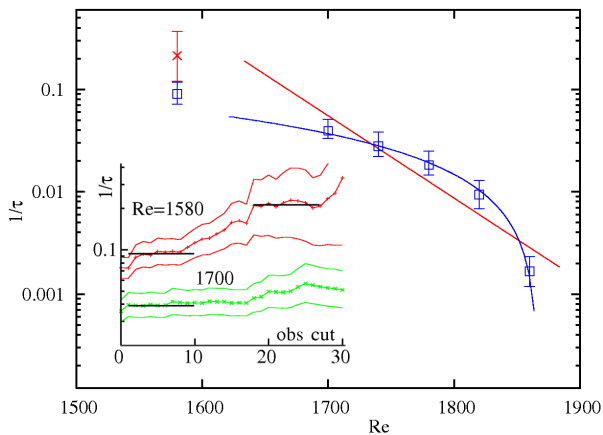, angle=0, width=80mm}
   \caption{\label{fig:tau1logfits} 
      Data with 95\% confidence intervals.  
      Inset:\ median and 95\% bracketing lines {\it vs.}\ number of observations removed. 
   }
\end{figure}
The data are affected by a transient time $t_0$ for adjustment from the initial condition
at $Re=1900$ to the operating $Re$.  This is of around $50\,D/U$ in the worst case.
In the inset to Fig.\ \ref{fig:tau1logfits} we calculate $1/\tau$ with 95\% confidence
intervals as a function of the number of smallest observations ignored.  
In the original analysis a maximum of
10\% of the data was removed and for $Re=1580$ gives the estimate indicated by the first solid bar. 
Cutting more leads to a rise in $1/\tau$ but the uncertainty becomes very large.
A possible time as suggested by Hof {\it et al.} is $\tau=4.9\,D/U$ 
(second higher bar, and red $\times$ on $Re$ plot), 
but it is an order of magnitude smaller than $t_0$.  As $\tau\ll t_0$ so few observations 
survive past the transient that nearly 20 of 40 observations must be cut. 
Aside from statistical uncertainty, this short time is far from the asymptotic
$\tau \to \infty$ and it is questionable that a self-sustaining mechanism
is really involved.  
The exponential fit of Hof {\it et al.}, however, relies heavily on this data point.

The largest step in $Re$ is from $1580$ to $1700$.  Here the adjustment in $Re$ from the initial condition
is smaller, the transient time is shorter, the typical observation time is longer
and the `tail' exponential distribution is clearly seen.  
It is this part of the distribution which has been fit --
after the first few points the estimate for $1/\tau$ (solid bar) is within the confidence interval 
even when large proportions of the data are ignored.

At $Re=1860$ Hof {\it et al.} state that the data is inconclusive because the probability
$P(T)$ is only shown for $T<1000\,D/U$
\endnote{
  In the supplementary to their article \cite{hof06}
  the numerical data (Fig.\ 2) is shown for
  $T<3000\,R/(2U) = 750\,D/U$, being less than our own limit.
}.
A maximum observation time is not a limitation provided that the probability 
distribution observed with in this time reflects that beyond it.  
The data fit a straight line on the log-plot and therefore
very well the exponential distribution, i.e.\ the process is memoryless.  
For the behaviour to change to a second memoryless process beyond $T=1000\,D/U$ is contradictory, 
requiring some property of the flow to be preserved to such long times. 

In Fig.\ \ref{fig:tau1logfits} the best-fit exponential and inverse scalings are fit
to data at the 5 largest $Re$.  It is evident that most points are significantly different at the 5\% level 
from the exponential scaling over this range of $Re$, no matter how it is fit.
In contrast $1/(Re_c-Re)$ with simple power $-1$ is a very natural fit.

\begin{acknowledgments}
   We thank David Leslie for advice on the bootstrapping method.
\end{acknowledgments}

\vspace{2pt}
\noindent
A.\ P.\ Willis and R.\ R.\ Kerswell,\\
\indent
Department of Mathematics,\\ 
\indent
University of Bristol, BS8 1TW, U.\ K.

\bibliography{reply_to_comment}

\end{document}